\documentclass[12pt]{iopart}
\usepackage{graphicx}
\def\be{\begin{equation}}
\def\ee{\end{equation}}
\def\bea{\begin{eqnarray}}
\def\eea{\end{eqnarray}}
\def\bdm{\begin{displaymath}}
\def\edm{\end{displaymath}}
\def\ba{\begin{array}}
\def\ea{\end{array}}

\begin{document}

\title[SUSY RLP vector for Dirac vortex in TI and graphene]
{Supersymmetric Runge-Lenz-Pauli vector for Dirac vortex in topological insulators and graphene}
\author{Chi-Ken~Lu and Igor F.~Herbut}
\address{Department of Physics, Simon Fraser University, Burnaby,
British Columbia V5A 1S6, Canada}

\ead{\mailto{chikenl@sfu.ca}}

\begin{abstract}

The Dirac mass-vortex at the surface of a topological insulator or
in graphene is considered. Within the linear approximation for the
vortex amplitude's radial dependence, the spectrum is a series of
degenerate bound states, which can be classified by a set of
accidental $SU(2)$ and supersymmetry generators (Herbut I F and Lu
C K 2011  {\it Phys. Rev.} B {\bf 83} 125412). Here we discuss
further the properties and manifestations of the supersymmetry of
the vortex Hamiltonian, and point out some interesting analogies
with the Runge-Lenz-Pauli vector in the non-relativistic hydrogen
atom. Symmetry breaking effects due to a finite chemical
potential, and the Zeeman field are also analyzed. We find that a
residual accidental degeneracy remains only in the special case of
equal magnitudes of both terms, whereas otherwise it becomes
removed entirely.

\end{abstract}


\maketitle

\vspace{10pt}

\section{Introduction}

Accidental degeneracy has been an interesting issue in
mathematical physics, with the classic examples ranging from
non-relativistic hydrogen atom to multi-dimensional harmonic
oscillator. The fact that states labeled by different quantum
numbers share the same energy is explained by identifying hidden
 constants of motion, unrelated to spatial symmetries that may be present in the problem.
 In hydrogen atom, for example, these are
the three Runge-Lenz-Pauli operators which, together with the
usual angular momentum, constitute a closed
$so(3)\times{so}(3)$ Lie algebra, and so imply the degeneracy among
all the levels of a given principle quantum
number.\cite{Jauch,Greenberg} For the hydrogen spectra in the
relativistic limit, however, the linear dispersion in the Dirac
Hamiltonian together with its matrix structure complicate the
degeneracy pattern and hinder the search for the hidden accidental constants of
motion.\cite{QMBook,relativisticHydrogen}

Recently, the interface between topologically trivial and non-trivial insulator
has become a topic of many investigations, because of its
hosting of the quasi-relativistic chiral Dirac
fermions.\cite{TIrev}  Depositing a type-II superconductor onto
the surface of a topological insulator, for example, could by the
proximity effect induce a mass-term for the Dirac fermions,
proportional to the superconducting order parameter.\cite{FuKane}
It is well known that, in general, in the vortex configuration of
such a mass-term, there exists a special topologically protected
state in the Dirac spectrum at zero energy.\cite{Jackiw1} Many
authors have recently elaborated on the properties of such
zero-modes in condensed matter systems, and derived several
unexpected consequences of their
existence.\cite{volovik,Ivanov,hou,herbut1,pi,wilczek,ghaemi,lehur,TeoKane}

Obtaining the entire bound spectrum for a general vortex profile
$\Delta(r)$, however, is typically not possible analytically, and
one has to resort to numerical methods. \cite{Kopnin}
Nevertheless, the corresponding Dirac Hamiltonian becomes
analytically solvable once the amplitude of the vortex is
linearized near the origin.\cite{Babak,IgorS}  In our previous
paper,\cite{IgorS} we showed that due to the presence of a closed
Clifford algebra associated with Dirac matrices the energy
spectrum even in this ``harmonic approximation" is non-trivial,
and, in particular, exhibits degeneracies due to the hidden and
accidental $SU(2)$ and supersymmetries. The eigenvalues are given
by $E=\pm\omega\sqrt{N}$, with an energy scale $\omega$, and with
integer $N$ corresponding to the total number of fictitious
fermions and bosons that can be formally defined in the problem,
and then distributed over two (in two dimensions) quantum states
in all possible ways. The associated degeneracy then comes out to
be $2N$, and when $N>1$ there are two distinct multiplets of the
effective ``angular momentum", with $j_>=N/2$ and $j_<=N/2-1$,
reminiscent of  the nonrelativistic spectrum of the hydrogen
atom.\cite{Jauch, Greenberg} Such a bound-state spectrum could be
observable by the scanning tunneling microscopy, and would
determine the specific heat of the vortex, for example.

In this paper, we investigate further the properties and the
algebra  of the supersymmetry generators, which are responsible
for the degeneracy between different multiplets of the angular
momentum. These generators commute with the square of the Dirac
vortex Hamiltonian, and their emergence in the problem  is closely
related to the existence of different square roots of the identity
matrix. We also establish a partial analogy with the
Runge-Lenz-Pauli vector for the text-book hydrogen
atom.\cite{Jauch, Greenberg} Besides its relevance to topological
insulators, our study is also pertinent to the insulating or
superconducting graphene, where the low-energy excitations are
also Dirac fermions, and the mass-gap can in principle be
generated dynamically, as well as externally .\cite{hou, herbut1,
ghaemi}

In addition to the usual symmetry of the Dirac spectrum around zero energy,
the chiral symmetry of the
Dirac Hamiltonian is also responsible for the appearance of the
linear combination of
odd and even occupation numbers of the fictitious fermions in the
bound-state wave functions. We use this property to
investigate the symmetry breaking effects of the chemical
potential and the Zeeman magnetic field, which in reality are present
both in graphene and in topological insulators. We find that the
degeneracy is completely lifted in general, but some residual
degeneracy remains in the special case when the two contributions are equal in
strength.

The paper has the following organization. In Sec. \ref{SUSY}, the
supersymmetry generators and their superalgebra are introduced in
terms of the associated supercharge. Sec. \ref{Matrix} is devoted
to explicitly constructing these symmetry generators. A comparison
with the Runge-Lenz-Pauli vector is also provided. The effects of
the finite chemical potential and the Zeeman field, and their connection
with the chiral operator are investigated in Sec. \ref{MuEffect}.
Finally, the discussion of validity of the linear approximation
made for the vortex amplitude, and comments on related problems in
the literature are offered in Sec. \ref{DisCon}. Some explicit
wave functions are exhibited in Sec. \ref{Appendix}.

\section{Emergent supersymmetry from $H^2$}\label{SUSY}

We define the  vortex configuration

\be
    \Delta(\vec{r})  = |\Delta(r)| ( \cos\theta, \sin\theta )
\ee  in the two-component order parameter $\Delta(\vec{r})$, which when uniform
provides a mass-gap for the two-dimensional Dirac fermions.
Such an order parameter may in principle arise from the
proximity-induced superconductivity in both graphene and
topological insulator, for example. $(r, \theta)$ are the usual polar coordinates in the plane.
The winding number
of the vortex is assumed to be unity. The spatial profile of the the order
parameter's amplitude will be taken to be  qualitatively similar
to the simple $|\Delta(r)|=\Delta_{\infty}\tanh{(r/\xi)}$, which
features the generic linear dependence on the radial coordinate
near the origin and the saturation to a finite value
$\Delta_{\infty}$ at infinity. $\xi$ is the characteristic
(coherence) length.

With the above qualifications in mind,
the linearized
Dirac Hamiltonian with the mass-vortex may be written as
\be
    H=c(\alpha_1 p_1+\alpha_2p_2)+\frac{\Delta_{\infty}}{\xi}(\beta_1x_1+\beta_2x_2)\:,\label{genericH}
\ee where the set of four $4\times{4}$ Hermitian matrix
$\{\alpha_i,\beta_i\}$, $i=1,2$, constitutes the
4-dimensional Clifford algebra. The fifth matrix $ \Gamma= \alpha_1\alpha_2\beta_1\beta_2 $ is responsible for
the chiral symmetry, since it evidently {\it anticommutes} with the Hamiltonian: $\{\Gamma,H\}=0$.

A supersymmetric representation of the Dirac Hamiltonian  in Eq.\ (\ref{genericH}) is possible both
because of the existence
of the standard bosonic representation for the operators of coordinates and momenta,

\be
    b_i=\sqrt{\frac{\Delta_{\infty}}{2c\xi}}x_i
    +i\sqrt{\frac{c\xi}{2\Delta_{\infty}}}p_i\:,
\ee  and of the fermionic representation for the \emph{four} Dirac matrices,\cite{IgorS}
\be
    a_i=\frac{\beta_i+i\alpha_i}{2}\:.
\ee Here $[b_i, b^\dagger _j] = \{ a_i, a^\dagger _j \}=
\delta_{ij}$, $[b_i, b_j] = \{ a_i, a_j \}= 0$, as usual. The two
transformations together yield the following supersymmetric
Hamiltonian,

\be
    H=\omega(a^{\dag}_1b_1+a^{\dag}_2b_2
    +a_1b^{\dag}_1+a_2b^{\dag}_2)\:,\label{H0}
\ee where the energy
$\omega=\sqrt{2c\Delta_{\infty}/\xi}$. Hereafter, $\omega$ will be assumed to be unity, and
will not be shown explicitly unless needed. The energy
eigenvalues in this representation become obvious,
 because
 \be
 H^2=\sum_{i=1,2}(a^{\dag}_ia_i+b^{\dag}_ib_i),
 \ee
turns out to be a simple sum of fermion and boson particle numbers.
Therefore, the energy levels can be labeled by the total
particle number $N$, and various ways of distributing
these fictitious particles according to their statistics leads to the degeneracy of $2N$.\cite{IgorS}
It is also easy to see that the zero-mode in this representation corresponds to the nondegenerate
vacuum $|0\rangle$, which is also an eigenstate of the chiral symmetry
operator $\Gamma$ with the eigenvalue of unity.

In order to see how the underlying supersymmetry emerges, we
first note the commutators $[H^2,a^{\dag}_i]=a^{\dag}_i$ and
$[H^2,b^{\dag}_i]=b^{\dag}_i$ for $i=1,2$. We then search for
other operators whose square has the
identical commutation relations with the particle operators.
 With the help of identity
$[A^2,X]=\{A,[A,X]\}=[A,\{A,X\}]$ for either $X=b^{\dag}_i$ or
$X=a^{\dag}_i$, one readily finds the following three such additional
supersymmetry operators:
\be
    A_i=\sum_{j,k=1,2}
    \sigma_{i,jk}(a^{\dag}_{j}b_{k}
    +b^{\dag}_{j}a_{k}), \label{vecA}
\ee which satisfy $A_i^2=H^2$ (without the summation convention).
In the above, the index $i$ denotes the i-th Pauli matrix, whereas
the subscripts $(j,k)$ label the matrix elements. The existence of
these operators is linked to the fact that the two-by-two
identity matrix $I_2$ has four distinct square roots. Together
with $H$, the four supersymmetry operators may be decomposed into
the supercharges and their Hermitian conjugates,\cite{Zee}
$(H,\vec{A})=Q_{\mu}+Q^{\dag}_{\mu}$. The time-component $\mu=0$
corresponds to $H$ and the rest of components refers to $\vec{A}$.
The supercharge is defined as
\be Q_{\mu}=   \sum_{j,k=1,2 }  b^{\dag}_{j}(\sigma_{\mu})_{jk}a_{k}.
\ee

We now determine the algebra satisfied by the supersymmetry
operators. First, due to the quantum statistics associated with
the $a$'s and $b$'s, the supercharges are mutually anticommuting:
\be
    \{Q_{\mu},Q_{\nu}\}=0\:.
\ee Next, it is useful to invoke the anticommutator

\be
    \{a^{\dag}_ib_j,b^{\dag}_la_m\}=a^{\dag}_ia_m\delta_{jl}
    +b^{\dag}_lb_j\delta_{im}\:,\label{Basic0}
\ee between boson-fermion mixed operators to compute the
anticommutators between $Q$'s and $Q^{\dag}$'s,

\be
    \{Q_{\mu},Q^{\dag}_{\nu}\}=\sum_{j,k=1,2}\left(
    b^{\dag}_{j}(\sigma_{\mu}\sigma_{\nu})_{jk}b_{k}
    +a^{\dag}_{j}(\sigma_{\nu}\sigma_{\mu})_{jk}a_{k}
    \right)\:,\label{BasicI}
\ee which will be crucial in the subsequent generation of the
constants of motion. Eq.\ (\ref{BasicI}) leads to the following simple
superalgebra for the components of the operator $\vec{A}$:

\be
    \{A_i,A_j\}=2H^2\delta_{ij}\:.\label{S3}
\ee When the time-component $H$ is involved in Eq.\
(\ref{BasicI}), on the other hand, one obtains the generators of
the $SU(2)$ symmetry of the Dirac Hamiltonian as

\be
    \{H,\vec{A}\}=4\vec{J}\label{S4}\:,
\ee defined  as,\cite{IgorS}

\be
    J_i=\frac{1}{2}\sum_{j,k=1,2}
    \sigma_{i,jk}(a^{\dag}_{j}a_{k}
    +b^{\dag}_{j}b_{k})\:.\label{vecJ}
\ee From Eq.\ (\ref{S3}) and (\ref{S4}), we then see that $\vec{J}$
is a constant of motion, since
\be
[J_i,H]=\frac{1}{4} [\{H,A_i\},H]=0.
\ee

Next, the algebra between $J$'s and $A$'s may be displayed in terms of
commutators, which also make the vector nature of the latter under the $SU(2)$ symmetry manifest.
The following identity,

\be
    [X^{\dag}_iX_j,X^{\dag}_lX_m]=X^{\dag}_iX_m\delta_{jl}
    -X^{\dag}_lX_j\delta_{im}\:,\label{BasicII}
\ee for both fermion $X=a$ and boson $X=b$ can be used to derive
the standard $SU(2)$ commutator algebra between the angular momentum generators,
\be
    [J_i,J_j]=i\epsilon_{ijk}J_k\:.\label{S1}
\ee
From the commutator between distinct species,
\be
    [X^{\dag}_iY_j,Y^{\dag}_lY_m]=X^{\dag}_iY_m\delta_{jl}\:,\label{BasicIII}
\ee for $X=a(b)$ and $Y=b(a)$, on the other hand,
 one can show that  $\vec{A}$ is indeed a vector under the $SU(2)$ algebra,
 i. e.
\be
    [A_i,J_j]=i\epsilon_{ijk}A_k\:.\label{S2}
\ee

Eqs.\ (\ref{S3}), (\ref{S4}), (\ref{S1}) and
(\ref{S2}) constitute the superalgebra that is
behind the degeneracy of the spectrum of $H$. The reader should note that Eq.\
(\ref{S4}), (\ref{S1}) and (\ref{S2}) are mutually consistent:
this can be seen, for example, through replacing with
$J_1=\{H,A_1\}/4$ to show that $[J_1,J_2]=\{[A_1,J_2],H\}/4=iJ_3$.
In order to include the chiral symmetry operator $\Gamma$ and
complete the algebra, we note that
$\{\Gamma, a_i\}=[\Gamma, b_i]=0$, which yields the final
relations,

\bea
    \{\Gamma,H\}=\{\Gamma,\vec{A}\}=0,\label{chiral}\\\:
    [\Gamma,\vec{J}]=0\:.
\eea

Let us remark in passing that the above derivation can be extended
to higher dimensions, to consider, for example, the
three-component mass hedgehog in the three-dimensional Dirac
equation.\cite{TeoKane,IgorS,Ryu} This case would involve
distributing fermions and bosons over three quantum states to
obtain the spectral degeneracies. The essential modification would
then be the replacement of the  matrices $\{I_2,\vec{\sigma}\}$
which appear in Eq.\ (\ref{vecA}) and (\ref{vecJ}) with another
set, consisting of an identity matrix and the eight Gell-Mann
matrices.\cite{Georgi} Again, the equalities Eq.\ (\ref{Basic0}),
(\ref{BasicII}), and (\ref{BasicIII}) are still valid between
three boson and three fermion operators. Therefore, the
commutation relations in Eq.\ (\ref{S1}) and (\ref{S2}) still
hold, except that the structure factor associated with $SU(3)$
algebra should be involved. Although $[A_i^2,H^2]=0$ also holds
here, $A_i^2$ does not coincide with $H^2$ as in $SU(2)$ case.
Nevertheless, the square of Hamiltonian still can be expressed as
the sum $H^2=(3/32)\sum_{i=1}^8A_i^2$.


\section{Representation of supersymmetry operators
 and connection with Runge-Lenz-Pauli vector}\label{Matrix}

The degeneracy pattern of the spectrum of the Hamiltonian in Eq.\
(\ref{genericH}) (Fig.1 in Ref.\cite{IgorS}) is similar to that
of the bound spectrum in hydrogen. It is therefore of interest to
investigate how the states within each eigenspace with $E=\sqrt{N}$
are connected to each other via the application of the operators
$\vec{J}$ and $\vec{A}$. In addition, since the
subspace of the second excited level with $N=2$ has a one-to-one
correspondence with that of the first excited level with $n=2$ in
the hydrogen atom, we should be able to see the exact correspondence
in algebraic relations between the two cases.

First, a matrix representation for vector operator $\vec{A}$ of
Eq.\ (\ref{vecA}) will be constructed. We work in the basis of
eigenstates $|E,j,m\rangle$ labelled by the respective eigenvalues
of $H$, $J^2$ and $J_3$. We recall that for $N>1$ there are two
allowed magnitudes of angular momentum $j_>=N/2$ and $j_<=j_>-1$
within the degenerate subspace with $E=\sqrt{N}$.\cite{IgorS} The
resultant degeneracy is then $2N$. The chiral operator $\Gamma$
also connects the states with opposite signs of the energy, as
usual.

According to Eq.\ (\ref{S2}), $A_3$ commutes with $J_3$ and $H^2$.
When $N>1$, it suffices then to consider the subset consisting of
the four states
$\{|{\pm}E,j_{>},m\rangle,|{\pm}E,j_{<},m\rangle\}$. We now
show that:
\be
    A_3|E,j_{>},m\rangle=\frac{2m}{E}|E,j_{>},m\rangle
    +\sqrt{E^2-\frac{4m^2}{E^2}}|-E,j_{<},m\rangle\:.\label{Key1}
\ee First, bracketing $\{H,A_3\}=4J_3$ between a pair of states
leads to

\be
    \langle{E},j_{>},m|A_3|E,j_{>},m\rangle=\frac{2m}{E},
\ee and to the conclusion that $A_3|E,j_{>},m\rangle$ is
orthogonal to $|E,j_{<},m\rangle$. Next, due to Eq.\
(\ref{chiral}), the matrix element

\be
    \langle{-E},j_{>},m|A_3|E,j_{>},m\rangle=-\langle{-E},j_{>},m|A_3|E,j_{>},m\rangle^{*}
\ee must be purely imaginary, or zero. Since $A_3$ in Eq.\
(\ref{vecA}) has only real coefficients, one concludes that
$A_3|E,j_{>},m\rangle$ is orthogonal to $|-E,j_{>},m\rangle$ as
well. Consequently, \be A_3|E,j_{>},m\rangle=\frac{2m}{E}
|E,j_{>},m\rangle+d|-E,j_{<},m\rangle \ee with a yet  undetermined
coefficient $d$. Similarly,

\be
    A_3|-E,j_{<},m\rangle=-\frac{2m}{E}|-E,j_{<},m\rangle+d|E,j_{>},m\rangle.
\ee Using then $A_3^2=H^2$ fixes the value of the coefficient $d$
to $d^2=E^2-4m^2/E^2$, as in Eq.\ (\ref{Key1}). Utilizing the
commutators $[A_3,J_{\pm}]=\pm{A}_{\pm}$ would then further
provide the matrix forms of $A_1$ and $A_2$, for example.



A compact representation for the operators $\vec{A}$ is available within the
degenerate subspace of dimension $4N$ and corresponding to $E^2=N$, which is the orthogonal sum
of the two energy eigenspaces with the positive and negative energies of equal magnitude.
It is straightforward
to write down the operators $H$ and $\vec{J}$ as block-diagonal matrices,
\bea
    \frac{H}{\sqrt{N}}&=&\sigma_3\otimes{I}_{2N}\:,\\
    J_i&=&I_2\otimes{L_i}\:.
\eea The identity matrix ${I}_{2N}$ and the projected (reducible)
angular momentum matrices $\vec{L}$ act within the degenerate
subspace of $E=\sqrt{N}$,  and of $E=-\sqrt{N}$. The chiral
symmetry operator $\Gamma$ that has the effect of reversing the
sign of energy can be similarly expressed as \be
    \Gamma=\sigma_1\otimes{I}_{2N}\:.
\ee
The vector
operator $\vec{A}$ projected onto the $E^2 =N$ subspace may be seen to possess the form,
\be
    A_i=\frac{2}{\sqrt{N}}\sigma_3\otimes{L_i}+2\sqrt{N}\sigma_2\otimes{K_i}\:,\label{RLPvector}
\ee
in which through Eq.\ (\ref{S4}) the first term
generates the above $\vec{J}$. The second term contains yet
another vector $\vec{K}$, which due to the multiplication by $\sigma_2$
cancels out in Eq.\ (\ref{S4}), but will be important for the analogy with the Runge-Lenz-Pauli vector
to be drawn shortly. Via Eq.\ (\ref{S2}) we then find
\be
    [K_i,L_j]=i\epsilon_{ijk}K_k\:,\label{ComuKL}
\ee
which resembles the commutation relation between the components of the
Runge-Lenz-Pauli vector and the orbital angular momentum in the  hydrogen atom.\cite{Greenberg}
Moreover, the coincidence between $H^2$ and $A_i^2$ implies the
following identity,

\be
    \frac{1}{N^2}L_i^2+K_i^2=\frac{1}{4}I_{2N}\:,\label{SumKL}
\ee to be true for all $i=1,2,3$. Consequently, the sum of squares
for both vectors turns into $L^2/N^2+K^2=3/4$. In the hydrogen's
bound spectrum, on the other hand, the corresponding identity reads
$K^2+L^2/n^2=1-1/n^2$,\cite{Greenberg} where the principle quantum
number is denoted by $n$. The two identities indeed match, but only
 when the respective degenerate subspaces contain \emph{two}
angular momentum multiplets. In hydrogen this is only true for the energy eigenspace with $n=2$,
which is  $2s\oplus{2p}$. In addition, one can also
rewrite the anticommutators in Eq.\ (\ref{S3}) in terms of the
vectors $\vec{L}$ and $\vec{K}$ as,
\be
    \{A_i,A_j\}=I_2\otimes\left(
    \frac{4}{N}\{L_i,L_j\}+4N\{K_i,K_j\}
    \right)\:,
\ee from which the pair of anticommutators adds up to zero for
different indices. For example,
$\{K_1,K_2\}=(-i)\{[K_2,L_3],K_2\}=(i/N^2)\{[L_2,L_3],L_2\}$ is
obtained through utilizing Eq.\ (\ref{ComuKL}) and (\ref{SumKL}),
and therefore the above sum of anticommutators vanishes.






\section{Effects of chemical potential and Zeeman field}\label{MuEffect}

In this section we study  the effects of chemical potential and of the Zeeman
coupling to the external magnetic field perpendicular to the surface of the
topological insulator. At  the surface of topological
insulator, the spin-rotational invariance is broken by the kinetic
energy term in Eq.\ (\ref{genericH}), which suggests that the
Hermitian matrix $i\alpha_1\alpha_2$ should correspond to the
out-of-plane component of the spin. Since the two mass-terms are rotated into each other
by the matrix $i\beta_1\beta_2$, we recognize this operator as representing the particle number,
i. e. the generator of the $U(1)$ symmetry associated with the superconducting phase.
 Therefore, the out-of-plane Zeeman field $h$ and chemical
potential $\mu$ shall enter the Hamiltonian through the couplings to
$i\alpha_1\alpha_2$ and $i\beta_1\beta_2$, respectively.
It will be useful, however, to represent these two terms by their sum $\Psi$
and their difference $\Pi$ instead:
\be
    \Psi\equiv\frac{i}{2}(\beta_1\beta_2+\alpha_1\alpha_2)=i(a_1^{\dag}a_2-a_2^{\dag}a_1)\:,
\ee
and
\be
    \Pi\equiv\frac{i}{2}(\beta_1\beta_2-\alpha_1\alpha_2)=i(a_1^{\dag}a_2^{\dag}+a_1a_2)\:.\label{Pi}
\ee We notice that the operator $\Psi$ coincides with the fermion
part of angular momentum operator $J_2$ in Eq.\ (\ref{vecJ}), so
$[\Psi,J_2]=[\Psi,H^2]=0$. The operator $\Pi$, on the other hand,
is a fermion ``pairing" term, which is also $SU(2)$ invariant:

\be
    [\Pi,\vec{J}]=0\:.
\ee


The chemical potential and the Zeeman coupling then can be written
together as $\varepsilon\Pi+\delta\Psi$ with $\varepsilon=\mu-h$
and $\delta=\mu+h$. Before proceeding further, we exploit the
$SU(2)$ invariance of the Hamiltonian in Eq.\ (\ref{genericH}) and
rotate for convenience the operator $\Psi$ into

\be
    \Psi=a^{\dag}_1a_1-a^{\dag}_2a_2\:,\label{Psi}
\ee which is the fermion piece of $J_3$.

Since $\Psi$ and $\Pi$ act on the states by changing only the
fermion part of wave functions, they can be viewed as two linear
and orthogonal operators within the fermion subspace
$\{|0\rangle,a^{\dag}_1|0\rangle,a^{\dag}_2|0\rangle,a^{\dag}_1a^{\dag}_2|0\rangle\}$,
in which the common, but for the present purposes irrelevant boson part is not shown explicitly.
$\Pi$ projects onto the subspace
$\{|0\rangle,a^{\dag}_1a^{\dag}_2|0\rangle\}$ consisting of even
number of fermions, since obviously
$\Pi{a}^{\dag}_{1(2)}|0\rangle=0$. Moreover,

\bea
    \Pi|0\rangle=ia^{\dag}_1a^{\dag}_2|0\rangle,\label{Pi1}\\\:
    {\Pi}a^{\dag}_1a^{\dag}_2|0\rangle=-i|0\rangle\:,
\eea
so $\Pi$ acts like the $\sigma_2$ within this subspace. $\Psi$, in contrast,
is a projection operator onto the subspace
$\{a^{\dag}_1|0\rangle,a^{\dag}_2|0\rangle\}$ consisting of only
one fermion:
\bea
    {\Psi}a^{\dag}_1|0\rangle=a^{\dag}_1|0\rangle,\\\:
    {\Psi}a^{\dag}_2|0\rangle=-a^{\dag}_2|0\rangle,
\eea
$\Psi$ therefore acts  analogously to $\sigma_3$ in this subspace.

We may also introduce the  projection operators from the chiral symmetry
operator $\Gamma$, namely,

\begin{displaymath}
   {\cal{P}}_{\pm}=(1\pm\Gamma)/2
\end{displaymath} because $\Gamma$ anticommutes with both fermion operators
$a_{1,2}$. This can be seen, for example, from the fact that
${\cal{P}}_{+}a^{\dag}_1|0\rangle=0$ and
${\cal{P}}_{-}a^{\dag}_1|0\rangle=a^{\dag}_1|0\rangle$ since the
zero-mode has $\Gamma|0\rangle=|0\rangle$. It follows that
$\Psi{{\cal{P}}_{+}}=\Pi{{\cal{P}}_{-}}=0$, which can also be seen
from the direct computation that yields

\be
    \Pi^2={{\cal{P}}_{+}},\ \Psi^2={{\cal{P}}_{-}}.
\ee The full Hamiltonian including the perturbations can therefore
be written as, \be
    H'=H
    +{{\cal{P}}_{+}}(\varepsilon\Pi){\cal{P}}_{+}+
    {\cal{P}}_{-}(\delta\Psi){\cal{P}}_{-}\:.\label{perturbH}
\ee

\begin{figure}
\input{epsf}
\includegraphics[scale=0.75]{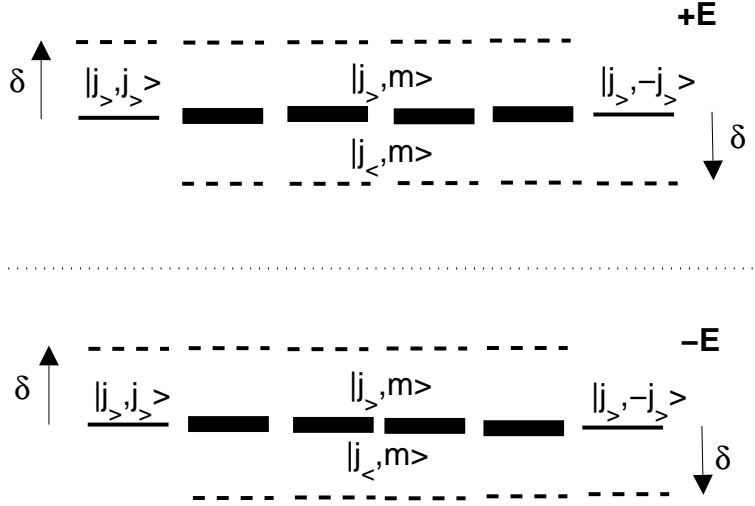}
\caption{The splitting of the degenerate multiplets $|E,j_{>},m\rangle$ and $|E,j_{<},m\rangle$
due to the effects of $\Psi$ in Eq.\ (\ref{perturbH}), for $N>1$ and $\mu=h=\delta/2$.
The thick solid lines stand for the two states with
the same quantum number $m$ but distinct $j$, which are coincident in energy for $\delta=0$.
The thin solid lines denote the unique states of maximal value of $|m|$
within this degenerate subspace. Dashed line are the shifted energies when $\delta\neq 0$.
The middle dotted line stands for the unperturbed zero energy state.}\label{perturbation1}
\end{figure}

It is easier to consider the effects of $\Pi$ and $\Psi$
separately. Since it commutes with $J_3$ and $H^2$, the state
$|E,j,m\rangle$ is by $\Psi$ coupled only to the state with the
same $m$ and the same $E^2$. As far as the coupling matrix for
$\Psi$ in Eq.\ (\ref{Psi}) is concerned, the only relevant states
are $|1(3)\rangle\equiv|\pm{E},j_{>},m\rangle$ and
$|2(4)\rangle\equiv|\pm{E},j_{<},m\rangle$ when $|m|<j_{>}$. On
the other hand, in the case of largest $|m|=j_{>}$, the relevant
states only include $|1\rangle$ and $|3\rangle$. It will be shown
shortly that $H+\delta\Psi$ is in this basis reduced to the
universal matrix, \be
    E\sigma_3\otimes{I_2}+\delta(I_2-\sigma_1)\otimes\sigma_3\:,\label{PsiForm}
\ee
when $|m|<j_{>}$. The two-dimensional matrix,
\be
    E\sigma_3\pm\delta(I_2-\sigma_1)\:,\label{PsiForm1}
\ee
on the other hand, is responsible for the cases of $m=\pm{j}_{>}$. In either case,
the universal matrix is independent of $m$, and the eigenvalues are
$\pm\sqrt{E^2+\delta^2}\pm\delta$. To the first order
of $\delta$, the splitting of the spectrum's degeneracy  is shown in
Fig.\ref{perturbation1}.


Let us explain the appearance of the above  universal matrix in more
detail. First consider $|m|<j_{>}$. Recall that $\Psi$
acts only on the one-fermion part of wave functions.
Hence, only the projected wave functions,

\be
    {{\cal{P}}_{-}}|\pm{E},j,m\rangle=\pm\left[c_{1(2)}f_{m}(b^{\dag}_1,b^{\dag}_2)
    a^{\dag}_1{\pm}c_{2(1)}g_{m}(b^{\dag}_1,b^{\dag}_2)a^{\dag}_2
    \right]|0\rangle\:,\label{projected}
\ee are needed when considering the effective couplings. The above
$f_m$ and $g_m$ are some general products of boson operators
$b^{\dag}$'s that will not affect the following analysis. Within
the bracket, the order of coefficients $c_1$ and $c_2$ as well as
$\pm$ sign is to denote the two orthogonal states corresponding to
$j=j_{>}$ and $j=j_{<}$, respectively, whereas the $\pm$ sign
preceding the bracket is to denote the sign of $E$. Now it is
straightforward to show that $\Psi$ is given by,

\be
    \delta\Psi=\delta(I_2-\sigma_1){\otimes}S\:,
\ee in which the minus sign preceding $\sigma_1$ appears because
opposite signs are associated with opposite-energy states in Eq.\
(\ref{projected}). The 2-by-2 matrix $S$ is associated with
coupling between $|1\rangle$ and $|2\rangle$. Explicitly,

\be
    S=(c_1^2-c_2^2)\sigma_3+2c_1c_2\sigma_1\:,\label{UniversalPsi}
\ee which yields a pair of eigenvalues $\pm(c^2_1+c^2_2)$.
We note that $c^2_1+c^2_2=1$ is dictated by the orthogonality
between wave functions in Eq.\ (\ref{projected}). Moreover, we can
rotate the space spanned by $|1\rangle$ and $|2\rangle$ so that
$S$ turns into a diagonal $\sigma_3$, which finally leads to the
matrix form in Eq.\ (\ref{PsiForm}). Similar method can be applied
to $|m|=j_{>}$ to obtain Eq.\ (\ref{PsiForm1}).

\begin{figure}
\input{epsf}
\includegraphics[scale=0.75]{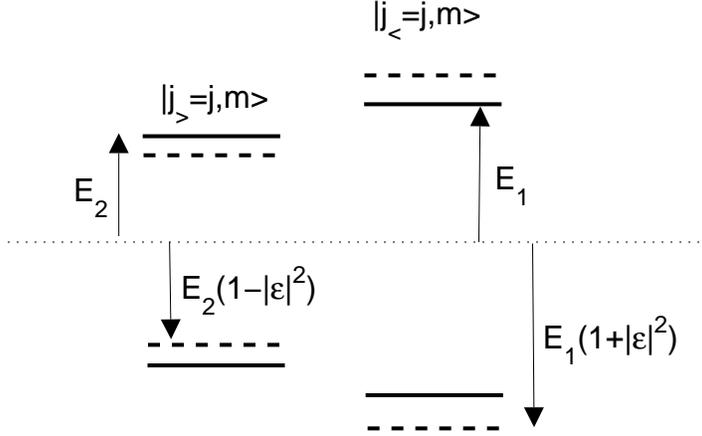}
\caption{The splitting between the states with the same quantum namber
$j$ but with different $E$, due to the $\Pi$-term in Eq.\
(\ref{perturbH}). This corresponds to $\mu=-h=\varepsilon/2$. The
solid and dashed lines represent the multiplets before and after,
respectively, the extra term $\varepsilon\neq 0$ is turned on.
 The dotted line
in the middle stands for the unperturbed zero-energy state.
The degeneracy due to
the $SU(2)$ rotational symmetry remains intact in this case.
}\label{perturbation2}
\end{figure}

Next we consider the coupling matrix for $\Pi$ in Eq.\ (\ref{Pi})
and the ensuing splitting of the  spectrum. Since
$[\Pi,\vec{J}]=0$ and $[\Pi,H]\neq{0}$, it suffices to consider
another subspace consisting of
$|1(2)\rangle\equiv|\pm{E}_1,j,m\rangle$ and
$|3(4)\rangle\equiv|\pm{E}_2,j,m\rangle$ for a given $j$ with the
constraint $E_1^2=E_2^2+2$. Note that $\Pi$ will only act on the
even-fermion part of wave functions
${{\cal{P}}_{+}}|E_{1(2)},j_{<(>)},m\rangle$. Explicitly, the one
including both fermions
$a^{\dag}_1a^{\dag}_2u_m(b^{\dag}_1,b^{\dag}_2)|0\rangle$
corresponds to $(E,j)=(E_1,j_<)$, while the one without any
fermion $u_m(b^{\dag}_1,b^{\dag}_2)|0\rangle$ is responsible when
$(E,j)=(E_2,j_>)$. As can be seen in the Appendix, $u_m$
represents a general product of boson operators $b^{\dag}$'s,
which will neither affect the coupling matrix of $\Pi$. Then, the
four-dimensional matrix representing  $H+\varepsilon\Pi$ assumes
the following universal form, \be
    E_sI_2\otimes{\sigma_3}+E_d{\sigma_3}\otimes{\sigma_3}+\varepsilon\ \sigma_2\otimes
    (I_2+\sigma_1)\:,
\ee where the energy is defined through
$E_{s(d)}=(E_1\pm{E}_2)/2$. This matrix then yield two new pairs
of eigenvalues, $|E_1'|=|E_1|\sqrt{1+|\varepsilon|^2}$ and
$|E_2'|=|E_2|\sqrt{1-|\varepsilon|^2}$, respectively. As shown in
Fig. \ref{perturbation2}, the degeneracy between a pair of distinct
$j$ is removed, while the accidental $SU(2)$ degeneracy remains unresolved.

The effects of the perturbations under consideration on the zero-mode
can also be simply analyzed by using the projection operators.
Obviously, due to the projection into odd-fermion
subspace, the term proportional to $\delta\Psi$ in Eq.\
(\ref{perturbH}) by itself has no effect on the zero-mode.
The $\Pi$ term in Eq.\ (\ref{perturbH}) can only
couple the zero-mode to another pair of rotationally invariant
states, i.e. the pair of opposite-energy states of $N=2$ and
$j=0$. When the $\Pi$-term alone is present, the corresponding three-dimensional
matrix that connects the $j=0$  states is then found to always have a zero eigenvalue.
Within the linear approximation, we therefore conclude, the zero-mode is still present
in the spectrum in the presence of the chemical potential and the  Zeeman field, at the special point
$|\mu|=|h|$. This agrees with the more general analysis, \cite{HerbutLu}
which shows that in the presence of chiral-symmetry breaking terms such as $h$ and $\mu$,
the zero-mode is implied under very broad conditions by the existence of an antilinear operator that anticommutes
with the Dirac vortex Hamiltonian.

In presence of both terms $\Psi$ and $\Pi$,
the calculation becomes rather complicated, and we will not pursue it further here.
As long as the
perturbations have weak spatial variation, however, our treatment in terms
of fermion operators only is still applicable. Presence of both perturbation in Eq.\
(\ref{perturbH}) introduces couplings which are not closed
within any finite subset of bound states. For each value of the
quantum number $m$, there will be a corresponding matrix of
infinite dimension, representing the coupling among all the states
with the same $m$. In this case no degeneracy is expected to survive in the spectrum.


\section{Discussions and Conclusions}\label{DisCon}

The accidental symmetry and the emergent supersymmetry due to $\vec{J}$ and $\vec{A}$ both depend crucially on the
assumed linearity as well as on isotropy of the vortex amplitude. Including
the higher order terms in  $|\Delta(r)|$ would
introduce combinations of higher powers of boson
operators, which would be expected to  remove the accidental degeneracy.
The anisotropy in the linear terms, on the other hand, would result in an asymmetry to
the square of Hamiltonian
$H^2=\sum_{i=1,2}m_i(a^{\dag}_ia_i+b^{\dag}_ib_i$), \cite{IgorS}
and $m_1\neq{m_2}$. Following the arguments in Sec. \ref{SUSY}, it
is evident that finding out the square roots of $H^2$ other
than $H$ itself is then impossible. Basically, it is due to the fact
that the matrix $(m_1+m_2)I_2/2+(m_1-m_2)\sigma_3/2$ has only one
independent square root when the $m_1 \neq m_2$.

Terms in the Dirac Hamiltonian with the  linear dependence on the
coordinates may also occur in graphene, for example, in presence
of the magnetic field perpendicular to the plane. In this case a
rather different supersymmetry was found.\cite{Ezawa} Yet another
similar but maybe a better known case is the Dirac
oscillator\cite{DiracOsc} with the Hamiltonian \be
H_{os}=\alpha_i(p_i-i\Gamma{x}_i). \ee In this case the crucial
difference from our problem lies in that the four Dirac matrices
$\{\alpha_1,\alpha_2,i\alpha_1\Gamma,i\alpha_2\Gamma\}$ that
appear in $H_{os}$ do not close a Clifford algebra, i.e. not all
the pairs of matrices anticommute. The square of the Hamiltonian,
$H_{os}^2$,  contains a spin-orbit coupling $(1-L_3S_3)\Gamma$ in
addition to the common part of oscillator Hamiltonian as in our
$H^2$. Here, the angular momentum $L_3=x_1p_2-x_2p_1$ and spin
$S_3=i\alpha_1\alpha_2$ are defined. Compared to the present
problem, the extra term in $H_{os}^2$ can be any integer and
half-integer, whereas the extra term $\sum_{i=1,2}a^{\dag}_ia_i$
in our problem can have the values $\{0,1,2\}$ only. These subtle
differences ultimately lead to a rather different underlying
(super)symmetry and the ensuing degeneracies of the spectrum.

By solving the differential equation for the zero-mode's
wave-function of our linearized Hamiltonian one finds its radial
dependence to be $\exp{(-\Delta_{\infty}r^2/2c\xi)}$. Thus, the
linearization of the vortex profile is a good approximation when
the coherence length is long, $c\ll \Delta_{\infty}\xi$. Also, in
the case of strictly linear vortex amplitude,  strong external
Zeeman field is not able to remove the zero-mode from the
spectrum. This stands in contrast to the previously studied case
of an order parameter which saturates to a finite value far from
the vortex.\cite{HerbutLu,LuHerbut} The spinor part of zero-mode
can be deduced from the matrix equation
$a_1|0\rangle=a_2|0\rangle=0$. Using the previous notations in
Ref.\cite{LuHerbut}, the fermion operators have the explicit
matrix form,

\be
    a_1=\left(\begin{array}{cccc}&i\sigma_2&I_2\\
    &I_2&-i\sigma_2\end{array}\right),\>\>
    a_2=\left(\begin{array}{cccc}&-i\sigma_1&iI_2\\
    &-iI_2&i\sigma_1\end{array}\right)\:,
\ee which yields the column vector $|0\rangle\propto(1,0,0,1)^T$
for the zero-mode.

Lastly, the treatment of chemical potential and Zeeman field with
two projections $\Psi$ and $\Pi$ is also relevant to a recent
study on two-velocity Weyl fermions realized in two-dimensional
optical lattice.\cite{Malcolm} The effective low-energy
Hamiltonian for the spinless fermions can be written as

\be
    H_{W}=k_x\alpha_1+k_y\alpha_2+\eta(ik_x\Gamma\beta_1+ik_y\Gamma\beta_2)\:,
\ee which yields the eigenvalues $\pm(1\pm\eta)k$, and $\eta$ is a
number indicating the difference in ``speed of light". Actually
the above Weyl Hamiltonian can be expressed as a more compact form
as one notices that the sets of matrices
$U=\{\alpha_1,\alpha_2,i\alpha_1\alpha_2\}$ and
$V=\{i\Gamma\beta_1,i\Gamma\beta_2,i\beta_1\beta_2\}$ form two
commutating $su(2)$ Lie algebras. Thus a generic Hamiltonian of

\be
    H_g=\vec{u}\cdot\vec{U}+\vec{v}\cdot\vec{V}\:,
\ee with components $\vec{u}=(k_x,k_y,0)$ and
$\vec{v}=\eta(k_x,k_y,0)$, can be co-rotated into
$|\vec{u}|U_3+|\vec{v}|V_3$. Then it is easy to see that
$H_W=(1+\eta)k\Psi+(1-\eta)k\Pi$, which explains that the
faster/slower Weyl fermions in fact correspond to the
odd-fermion/even-fermion subspaces in our definition.

In conclusion, for the two-dimensional linearized vortex on the
superconducting surface of topological insulator, we defined the
corresponding supersymmetric Runge-Lenz-Pauli operator, and
discussed the connection to its counterpart in the hydrogen atom.
We also studied the effects of the physical perturbations such as
finite chemical potential and Zeeman coupling of the spin to the
out-of plane magnetic field on the spectrum, and showed how these
can be economically accounted for in terms of the even/odd-fermion
projection operators which we introduced. In the most general
case, the accidental degeneracy of the spectrum is removed by
these perturbations, but the energy of the zero-mode is unaltered.
Unexpectedly, however, we found that some of the accidental
degeneracy remains present when the chemical potential and the
Zeeman field have identical magnitudes.

\begin{center}
{\bf Acknowledgment}
\end{center}

This work has been supported by the NSC Taiwan (C.K.L.) and the NSERC
of Canada (I.F.H.).

\section{Appendix}\label{Appendix}

Within the subspace of states of $E^2=N$, the wave functions with
the maximum eigenvalue of $J_3$ belong to the $j_>$-multiplet.
Explicitly, the pair of opposite-energy states can be written as,

\be
    |E=\pm\sqrt{N},j_>,m=\frac{N}{2}\rangle=\frac{1}{\sqrt{2}}
    \left(
    \frac{(b^{\dag}_1)^N}{\sqrt{N!}}\pm\frac{a^{\dag}_1(b^{\dag}_1)^{N-1}}{\sqrt{(N-1)!}}
    \right)|0\rangle\:.
\ee Within the same multiplet, the states with one lower in $m$ is
obtained by applying $J_{-}$ to the above states. The even-fermion
part of $|E=\pm\sqrt{N},j_>,m=j_>-1\rangle$ is,

\be
    \frac{1}{\sqrt{2}}
    \frac{b^{\dag}_2(b^{\dag}_1)^{N-1}}{\sqrt{(N-1)!}}
    |0\rangle\:,
\ee while the associated odd-fermion part is,

\be
    \pm\frac{1}{\sqrt{2}}\left[
    \sqrt{\frac{N-1}{N}}
    \frac{a^{\dag}_1b^{\dag}_2(b^{\dag}_1)^{N-2}}{\sqrt{(N-2)!}}
    +\frac{1}{\sqrt{N}}
    \frac{a^{\dag}_2(b^{\dag}_1)^{N-1}}{\sqrt{(N-1)!}}
    \right]|0\rangle\:,
\ee where the above $\pm$ denotes the corresponding sign of
energy. In the other multiplet of $j_<$, the state
$|E=\pm\sqrt{N},j_<,m=j_>-1\rangle$ has the same $m$ as the above
one, and they are orthogonal to each other. Therefore, its
even-fermion part is given by,

\be
    \frac{a^{\dag}_1a^{\dag}_2(b^{\dag}_1)^{N-2}}{\sqrt{(N-2)!}}|0\rangle\:,
\ee while the associated odd-fermion part is,

\be
    \mp\left[
    \frac{1}{\sqrt{N}}\frac{a^{\dag}_1b^{\dag}_2(b^{\dag}_1)^{N-2}}{\sqrt{(N-2)!}}-
    \sqrt{\frac{N-1}{N}}\frac{a^{\dag}_2(b^{\dag}_1)^{N-1}}{\sqrt{(N-1)!}}
    \right]|0\rangle\:,
\ee in which a reverse sign is used to denote the corresponding
positive/negative energies in this multiplet. This choice is
necessary for being consistent with the choice of sign for the
unknown $d$ in Eq.\ (\ref{Key1}). One can find out the rest of
wave functions within this subspace by successively applying the
ladder operator $J_{-}$ to the above states in both multiplets.

\end{document}